\documentclass[superscriptaddress,preprintnumbers,showpacs,twocolumn]{revtex4}
\usepackage[tbtags]{amsmath}
\usepackage{amsfonts}
\usepackage{mathrsfs}
\usepackage{amssymb}
\usepackage{graphicx}
\usepackage{epsfig,graphicx,times}
\usepackage{subfigure}
\usepackage{color}
\usepackage{amssymb}
\usepackage{amsmath}
\usepackage{graphicx}
\usepackage{dcolumn}
\usepackage{bm}
\usepackage{float}
\usepackage[abs]{overpic}

\begin{document}

\title{Single-photon transport in a one dimentional waveguide coupling to a hybrid atom-optomechanical system}
\author{W. Z. Jia}
\affiliation{Department of Physics and
Center of Theoretical and Computational Physics, The University of
Hong Kong, Pokfulam Road, Hong Kong, China}
\author{Z. D. Wang}
\email{zwang@hku.hk}
\affiliation{Department of Physics and
Center of Theoretical and Computational Physics, The University of
Hong Kong, Pokfulam Road, Hong Kong, China}

\begin{abstract}
We explore theoretically the single-photon transport in a single-mode waveguide
that is coupled to a hybrid atom-optomechanical system in a strong
optomechanical coupling regime. Using a full quantum real-space approach,
transmission and reflection coefficients of the propagating
 single-photon in the waveguide are obtained. The influences
 of atom-cavity detuning and the dissipation of atom on the transport
 are also studied.  Intriguingly, the obtained spectral features can reveal the
 strong light-matter interaction in this hybrid system.
\end{abstract}

 \maketitle

\bigskip

\section{\label{1}INTRODUCTION}

\begin{figure}[t]
\centering \includegraphics[width=0.4\textwidth]{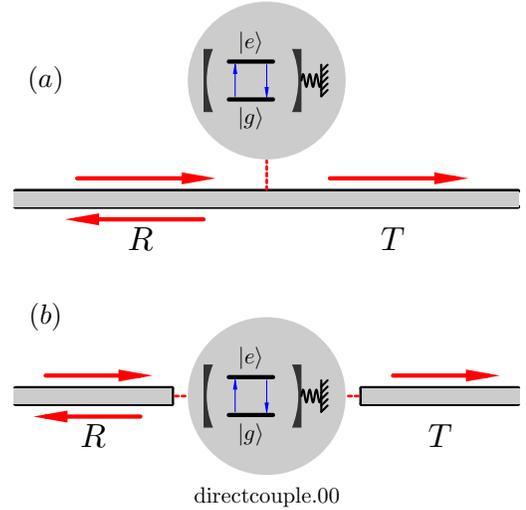} \centering \caption{ (Color
online) Schematic plot of a coupling system considered here. An optomechanical cavity
interacting with a two-level atom is coupled to a single-mode
waveguide, in which single photons propagate along the arrow direction. (a)
Side-coupled cases. (b) Direct coupled cases.}
\label{OMAsystem}
\end{figure}

Recently, research of controllable single photon transport in low dimensional
systems has attracted a growing interest for its significant
importance in quantum control including quantum information processing. Usually, this kind of
manipulation is achieved by strongly coupling a propagating
single-photon in the waveguide to a local quantum
system~\cite{Shen2005,Shen2009,Zhou2008}. A desired single-photon
control process is resulted from an interference between the directly
transmitted photon and the photon re-emitted by the emitter.
Specifically, such a waveguide-emitter system can be realized by a
photonic nanowire with an embedded quantum dot~\cite{photonic},
surface plasmons coupled to a single two-level
emitter~\cite{plasmon}, a superconducting transmission line coupled
to a superconducting artificial atom~\cite{NECscience2010}, or a
single-mode waveguide coupled to a cavity interacting with a
two-level atom~\cite{cQED1,cQED2,cQED3,cQED4}.

As is known,  a new type of optomechanical cavity was also developed to
couple photons and phonons via radiation pressure. Significant research interest in this 
frontier of optomechanics is motivated by its potential
applications in ultrasensitive measurements, quantum information
processing, and implementation of novel quantum phenomena at macroscopic scales
~\cite{Kippenberg2008,Marquardt2009,Aspelmeyer2013}. 
Important experimental progress on optomechanical systems has recently been made to reach the so-called single-photon strong coupling regime~\cite{ExpStrongCoup1,ExpStrongCoup2,ExpStrongCoup3,ExpStrongCoup4,ExpStrongCoup5,ExpStrongCoup6,ExpStrongCoup7},
where the single-photon coupling strength of the radiation pressure
is comparable to (or even larger than) the cavity decay rate. This
progress also inspired a series of theoretical investigations,
including photon blockade~\cite{Rabl2011 ,Nunnenkamp2011,Liao2013}
and photon-induced tunneling~\cite{Liu2013}, single-photon
cooling~\cite{SPcooling}, optomechanically induced transparency in
the single-photon strong coupling regime~\cite{OMIT}, and
optomechanical instability~\cite{OM instability}. In most quantum
optomechanical devices, the cavity is side or direct coupled to a
waveguide~\cite{Aspelmeyer2013}. Thus in the
single-photon regime, optomechanical systems, rather than traditional quantum emitters, may enable us 
 to control the propagating
single-photon in the waveguide. Also, the single-photon transmission
spectra can be used to probe and characterize the strong-coupling
regime~\cite{Liao2012,Xiao2013}.

In this paper, we explore theoretically  the single photon transport in a
waveguide coupled to a hybrid atom-optomechanical system in
the single-photon strong coupling regime. This kind of hybrid
atom-optomechanical system consists of an optomechanical cavity
interacting with a single two-level atom, and is suggested to
achieve a strong coupling between a single trapped atom and the motion
of a membrane~\cite{OMwithAtom1}. Notably,  a weak continuous-wave laser scattering
problem in this hybrid atom-optomechanical system was perturbatively
treated in a recent study by assuming the weak coupling,~\cite{OMwithAtom2}. Here, we employ a full quantum-mechanical
approach~\cite{Shen2005,Shen2009,Xiao2013} to study the transmission
and reflection properties of the propagating photon in the waveguide
in the strong optomechanical coupling regime. Our results also show
that the single-photon transmission and reflection spectra can be used
to probe and characterize the strong light-matter interaction in
this kind of hybrid systems.

The paper is organized as follows. In Sec.~\ref{2}, we introduce our model
 for describing the single-photon transport. Then, in Sec.~\ref{3},
 we look into the single-photon transport properties in detail.
 The influences of detuning and dissipation are also addressed. Finally,
  further discussions and  conclusions are given in Sec.~\ref{4}.

\section{\label{2}HAMILTONIAN AND THE SOLUTIONS}

We consider a hybrid atom-optomechanical system (i.e., a single
two-level atom coupled to an optomechanical cavity) to be coupled to an
open one dimensional waveguide. With well-developed techniques for
confining a single atom in a usual optical cavity~\cite{Kimble2005}, it
seems achievable in the
near future to couple atoms with optomechanical cavities. Usually, the atom-optomechanical system can either 
side-coupled or directly coupled to a waveguide, which is schematically
illustrated in Fig.~\ref{OMAsystem}(a) and (b). In this paper, we
focus on the single-photon transport problem of the side-coupling
case, for one can straightforwardly map the reflection amplitude of the
side-coupled case into the transmission amplitude of the
direct-coupled cases~\cite{Shen2009}. A model Hamiltonian of this system
may be written as ($\hbar=1$)
\begin{widetext}
\begin{eqnarray}
\hat{H} &=&\int dxa_{R}^{\dagger}\left(x\right)\left(-iv_{g}\frac{\partial}{\partial x}\right)a_{R}\left(x\right)+\int dxa_{L}^{\dagger}\left(x\right)\left(iv_{g}\frac{\partial}{\partial x}\right)a_{L}\left(x\right)\nonumber \\
 && +\frac{\omega_{a}}{2}\sigma_{z}-i\gamma_{a}\left\vert e\right\rangle _{a}\left\langle e\right\vert _{a}+\omega_{c}c^{\dagger}c+\Omega b^{\dagger}b-g_{0}c^{\dagger}c\left(b+b^{\dagger}\right)+\lambda\left(c\sigma^{+}+c^{\dagger}\sigma^{-}\right)\nonumber \\
 && +V\int dx\delta\left(x\right)\left(a_{R}^{\dagger}\left(x\right)c+a_{R}\left(x\right)c^{\dagger}+a_{L}^{\dagger}\left(x\right)c+a_{L}\left(x\right)c^{\dagger}\right).
\end{eqnarray}
\end{widetext}
The first line denotes the waveguide optical mode, where $v_{g}$
is the group velocity of the photons, and
$a_{R}^{\dagger}\left(x\right)$ ($a_{L}^{\dagger}\left(x\right)$) is
a bosonic operator creating a right-going (left-going) photon at
$x$. The second line describes the isolated atom-optomechanical
system, where $c^{\dagger}$ ($b^{\dagger}$) is the photon (phonon)
creation operator, $\sigma^{+}$($\sigma^{-}$) is the atomic raising
(lowering) operator generating transition between ground state and
exited state: $\sigma^{+}\left\vert g\right\rangle _{a}=\left\vert
e\right\rangle _{a},$ $\sigma^{-}\left\vert e\right\rangle
_{a}=\left\vert g\right\rangle _{a}$. $\omega_{a}$ is the atomic
transition frequency, $\omega_{c}$ is the cavity resonance
frequency, $\Omega$ is the mechanical frequency, $g_{0}$ is the
single-photon coupling strength of the radiation pressure between
the cavity and the mirror, $\lambda$ is the coupling strength
between the cavity and the atom, $\gamma_{a}$ is the dissipation
rate of the the atom, due to coupling to the reservoir. The third
line represents the coupling between the waveguide and the
atom-optomechanical system, where $V$ is the coupling strength
between the cavity and the waveguide. And the according
cavity-waveguide's decay rate can be defined as
$\Gamma=V^{2}/v_{g}$ ~\cite{Shen2009}. Note that in our treatment, it is assumed that the
majority of the decayed light from the cavity is guided into
waveguide modes, i.e., the "strong coupling" exists between the
cavity and the waveguide~\cite{HXZheng2011}. Thus the decay rate
$\kappa$ of the cavity into channels other than the 1D continuum is
negligible. We also assume that the decay rate $\gamma_{M}$ of the
mirror motion is much smaller than the cavity-waveguide's decay
rate. As a result, $\Gamma\gg\kappa,\gamma_{M}$, the optomechanical
decoherence processes can safely be ignored.

For an input one-photon Fock state, the stationary state of the
system satisfies the eigen equation
\begin{equation}
H\left\vert \epsilon\right\rangle =\epsilon\left\vert
\epsilon\right\rangle .
\label{eigen eq}
\end{equation}
We assume that, initially, the mirror is in state
$\left|n_{0}\right\rangle _{b}$, the atom is in the ground state and
the cavity is empty, and a single-photon comes from the left
with energy $v_{g}k$ with $k$ as the wave vector of the photon. In this case, the total energy of the coupled
system is $\epsilon=-\omega_{a}/2+v_{g}k+n_{0}\Omega$. In
the single-photon subspace, $\left\vert \epsilon\right\rangle$ can
be expanded as
\begin{eqnarray}
\left\vert \epsilon\right\rangle  & =&\sum_{n}\int dx\varphi_{R}\left(x,n\right)a_{R}^{\dagger}\left(x\right)\left\vert \varnothing\right\rangle \left\vert n\right\rangle _{b}\nonumber\\
&&+\sum_{n}\int dx\varphi_{L}\left(x,n\right)a_{L}^{\dagger}\left(x\right)\left\vert \varnothing\right\rangle \left\vert n\right\rangle _{b}\nonumber \\
 & &+\sum_{n}e_{n}c^{\dagger}\left\vert \varnothing\right\rangle \left\vert \tilde{n}\right\rangle _{b}+\sum_{n}f_{n}\sigma^{+}\left\vert \varnothing\right\rangle \left\vert n\right\rangle _{b}\label{eigen state},
\end{eqnarray}
where $\left\vert \varnothing\right\rangle =\left\vert
0\right\rangle _{k}\left\vert 0\right\rangle _{c}\left\vert
g\right\rangle _{a}$ is the vacuum state, with zero photon in both
the waveguide and the cavity, and with the atom in the ground state.
$\left\vert n\right\rangle _{b}$ represents the number state of the
mechanical mode. $\varphi_{R,L}\left(x,n\right)$ is the
single-photon wave function in the R/L mode. $e_{n}$ and $f_{n}$ are
excitation amplitudes of the cavity and the atom, respectively.
$\left\vert \tilde{n}\right\rangle
_{b}=\exp\left[\frac{g_{0}}{\Omega}\left(b^{\dagger}-b\right)\right]\left\vert
n\right\rangle _{b}$ is the single-photon displaced number state of
the mechanical oscillator satisfying the eigen equation
\begin{eqnarray}
\left[\omega_{c}c^{\dagger}c+\Omega
b^{\dagger}b-g_{0}c^{\dagger}c\left(b+b^{\dagger}\right)\right]\left\vert
1\right\rangle _{c}\left\vert \tilde{n}\right\rangle
_{b}
\nonumber\\
=\left(\omega_{c}+n\Omega-\delta\right)\left\vert 1\right\rangle
_{c}\left\vert \tilde{n}\right\rangle _{b},
\end{eqnarray}
where $\delta=g_{0}^{2}/\Omega$ is the
photon-state frequency shift caused by a single-photon radiation
pressure.

By substituting Eq.~\eqref{eigen state} into Eq.~\eqref{eigen eq}, we obtain the following
equations of motion
\begin{subequations}
\begin{eqnarray}
-iv_{g}\frac{\partial\varphi_{R}\left(x,n\right)}{\partial x}+\delta\left(x\right)V\sum_{m}e_{m}U_{nm}\nonumber\\
=\left(\epsilon+\frac{\omega_{a}}{2}-n\Omega\right)\varphi_{R}\left(x,n\right),\label{Eq of M1}
\\
iv_{g}\frac{\partial\varphi_{L}\left(x,n\right)}{\partial x}+\delta\left(x\right)V\sum_{m}e_{m}U_{nm}\nonumber\\
=\left(\epsilon+\frac{\omega_{a}}{2}-n\Omega\right)\varphi_{L}\left(x,n\right),\label{Eq of M2}
\\
V\int dx\delta\left(x\right)\left[\varphi_{R}\left(x,n\right)+\varphi_{L}\left(x,n\right)\right]+\lambda f_{n}\nonumber\\
=\sum_{m}\left(\epsilon+\frac{\omega_{a}}{2}-\omega_{c}-m\Omega+\delta\right)e_{m}U_{nm},\label{Eq of M3}
\\
\lambda\sum_{m}e_{m}U_{nm}=\left(\epsilon-\frac{\omega_{a}}{2}-n\Omega+i\gamma_{a}\right)f_{n},\label{Eq of M4}
\end{eqnarray}
\end{subequations}
with $U_{nm}=\left\langle n\mid\tilde{m}\right\rangle
_{b}$.

Assuming that the mirror is initially prepared in state
$\left|n_{0}\right\rangle _{b}$ and a single-photon comes from
the left with energy $v_{g}k$,
$\varphi_{R}\left(x,n\right)$ and
$\varphi_{L}\left(x,n\right)$ should take the form
\begin{subequations}
\begin{eqnarray}
\varphi_{R}\left(x,n\right) & = & \theta\left(-x\right)\delta_{nn_{0}}e^{i\left(k+\left(n_{0}-n\right)\frac{\Omega}{v_{g}}\right)x}\nonumber\\
&&+\theta\left(x\right)t_{n}e^{i\left(k+\left(n_{0}-n\right)\frac{\Omega}{v_{g}}\right)x},
\label{ansatz1}\\
\varphi_{L}\left(x,n\right) & = &
\theta\left(-x\right)r_{n}e^{-i\left(k+\left(n_{0}-n\right)\frac{\Omega}{v_{g}}\right)x}\label{ansatz2},
\end{eqnarray}
\end{subequations}
where $t_{n}$ and $r_{n}$ are the transmission and reflection
amplitude, respectively. Substituting Eqs.~\eqref{ansatz1} and
~\eqref{ansatz2} into Eqs.~\eqref{Eq of M1}-\eqref{Eq of M4}, 
the equations for $t_{n}$, $r_{n}$, $e_{n}$ and $f_{n}$ are given by
\begin{subequations}
\begin{eqnarray}
-iv_{g}\left(-\delta_{nn_{0}}+t_{n}\right)+V\sum_{m}e_{m}U_{nm}=0,
\label{aa}\\
-iv_{g}r_{n}+V\sum_{m}e_{m}U_{nm}=0,
\label{bb}\\
\frac{1}{2}V\left[\delta_{nn_{0}}+t_{n}+r_{n}\right]+\lambda f_{n}\nonumber\\
=\sum_{m}\left(\Delta_{c}+\left(n_{0}-m\right)\Omega+\delta\right)e_{m}U_{nm},
\label{cc}\\
\lambda\sum_{m}e_{m}U_{nm}=\left(\Delta_{c}-\Delta_{ac}+\left(n_{0}-n\right)\Omega+i\gamma_{a}\right)f_{n},
\label{dd}
\end{eqnarray}
\end{subequations}
with $\Delta_{c}=v_{g}k-\omega_{c}$,
$\Delta_{ac}=\omega_{a}-\omega_{c}$. If
$\lambda\ll\Gamma,\gamma_{a}$, we can have the series solutions of
$r_{n}$ and $t_{n}$:
\begin{subequations}
\begin{eqnarray}
&&r_{n}  =-i\Gamma\left(\sum_{n'}\frac{ U_{nn'}U_{n_{0}n'}^{*}}{\tilde{\Delta}_{c}\left(n'\right)}
\right.
\nonumber
\\
&&\left.+\sum_{n'mn''}\frac{\lambda^{2}U_{nn'}U_{mn'}^{*}U_{mn''}U_{n_{0}n''}^{*}}{\tilde{\Delta}_{c}\left(n'\right)\tilde{\Delta}_{a}\left(m\right)\tilde{\Delta}_{c}\left(n''\right)}\right.\nonumber
\\
&& \left.+\sum_{n'mn''m'n'''}\frac{\lambda^{4}U_{nn'}U_{mn'}^{*}U_{mn''}U_{m'n''}^{*}U_{m'n'''}U_{n_{0}n'''}^{*}}{\tilde{\Delta}_{c}\left(n'\right)\tilde{\Delta}_{a}\left(m\right)\tilde{\Delta}_{c}\left(n''\right)
\tilde{\Delta}_{a}\left(m'\right)\tilde{\Delta}_{c}\left(n'''\right)}\right.
\nonumber
\\
&&\left.
+\cdots\right),
\\
&&t_{n}=\delta_{nn_{0}}+r_{n}
\end{eqnarray}
\end{subequations}
with
\begin{eqnarray}
\tilde{\Delta}_{c}\left(m\right)&=&\Delta_{c}+\left(n_{0}-m\right)\Omega+\delta+i\Gamma,
\nonumber
\\
\tilde{\Delta}_{a}\left(n\right)&=&\Delta_{c}-\Delta_{ac}+\left(n_{0}-n\right)\Omega+i\gamma_{a}.
\nonumber
\end{eqnarray}
For an arbitrary $\lambda$, Eqs.~\eqref{aa}-\eqref{dd} can be
solved numerically by choosing the upper limit of $n$ large enough,
namely, $n_{max}\gg n_{0}$, and solving the attained
$4\left(n_{max}+1\right)$ equations. Note that $r_{n}$ ($t_{n}$)
represents the amplitude of reflecting (transmitting) a
single-photon with frequency $v_{g}k-\left(n-n_{0}\right)\Omega$.
Thus the total single-photon transmission and reflection
coefficients should be given by
\begin{equation}
T=\sum_{n}\left|t_{n}\right|^{2}, \,\, R=\sum_{n}\left|r_{n}\right|^{2}.\end{equation}

\section{\label{3}SINGLE-PHOTON SCATTERING SPECTRA}

\subsection{\label{A}Single-photon transmission(reflection) coefficient: atom
 cavity in tune and nondissipative case}

\begin{figure}[t]
\includegraphics[width=0.5\textwidth]{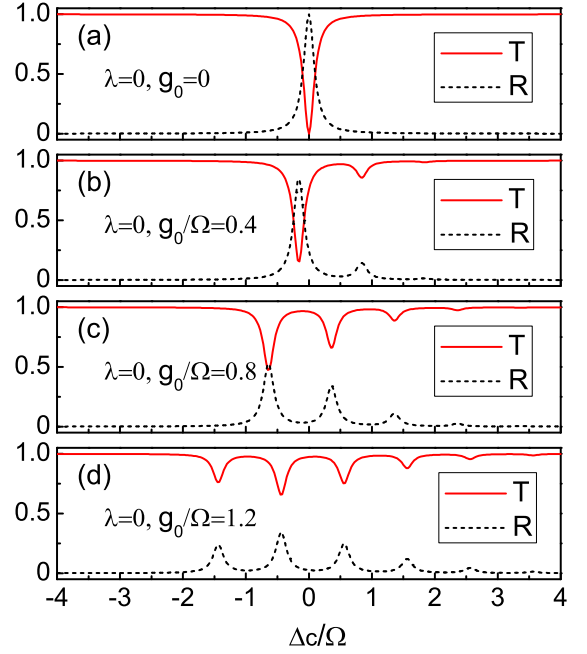}
\caption{ (Color online) Single-photon transmission (reflection) spectra of a standard optomechanical cavity (i.e., the cavity-atom coupling strength
$\lambda=0$) for various $g$. The cavity-waveguide decay rate
$\Gamma=0.1\Omega$ is chosen for plotting the spectra.}
\label{NoAtom}
\end{figure}

\begin{figure*}[t]
\includegraphics[width=0.8\textwidth]{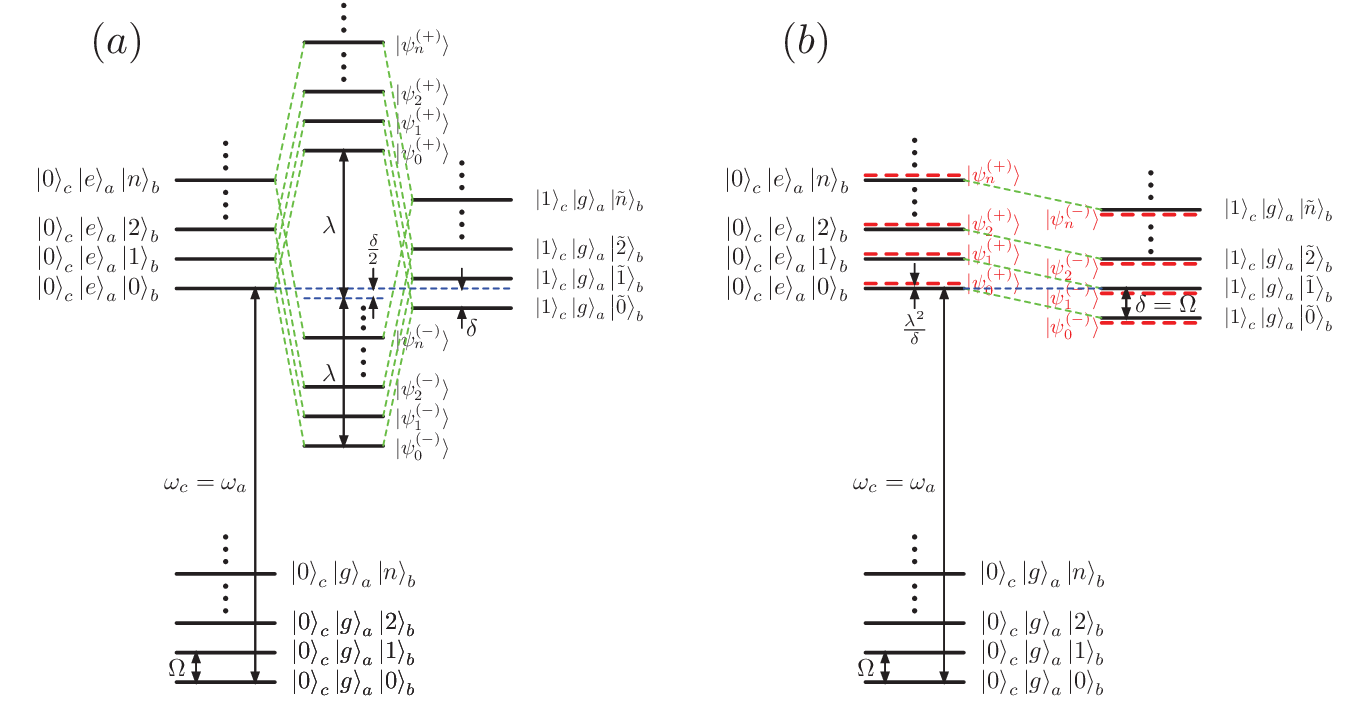}
\caption{ (Color online) The energy-level structure of a hybrid atom-optomechanical cavity (limited to the zero- and one-photon subspaces) in the single-photon strong coupling regime, where the two-level atom is in resonance
with the cavity (i.e., $\Delta_{ac}=0$).  The atom-cavity coupling
strength $\lambda\gg\Omega\gg\Gamma$ for (a); and $\lambda<\Gamma$  with  $g_{0}=\Omega$  (i.e., $\delta=\Omega$) for (b).}
\label{EL}
\end{figure*}

\begin{figure}[t]
\includegraphics[width=0.5\textwidth]{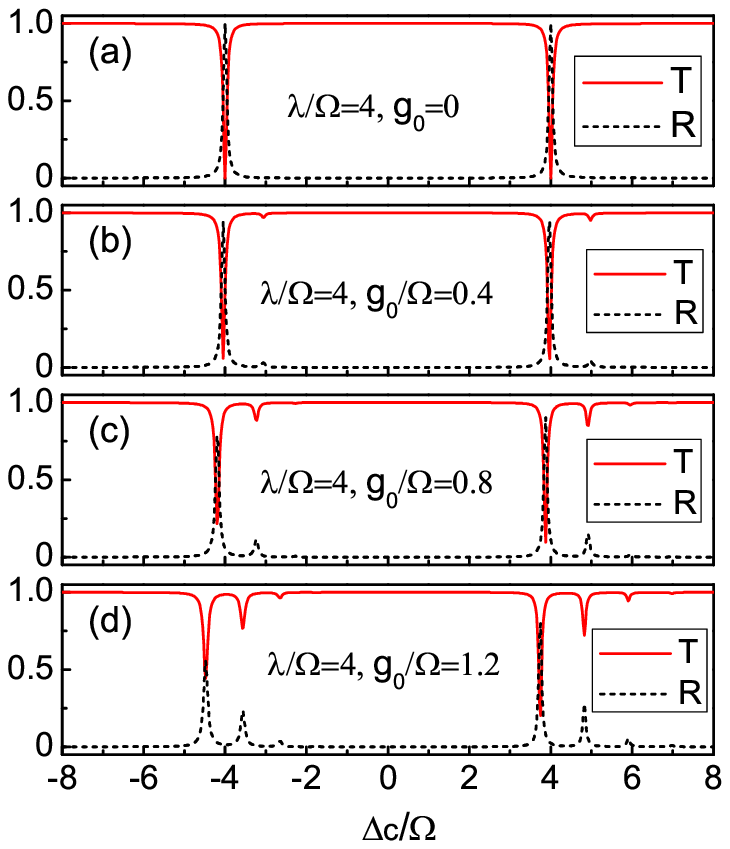}
\caption{(Color online) Single-photon transmission(reflection) spectra of the
atom-optomechanical system for $\lambda\gg\Gamma$. The
parameters are $\lambda=4\Omega$, $\Delta_{ac}=0$, $\gamma_{a}=0$,
and $\Gamma=0.1\Omega$.}
\label{Rabilike}
\end{figure}

\begin{figure}[t]
\includegraphics[width=0.5\textwidth]{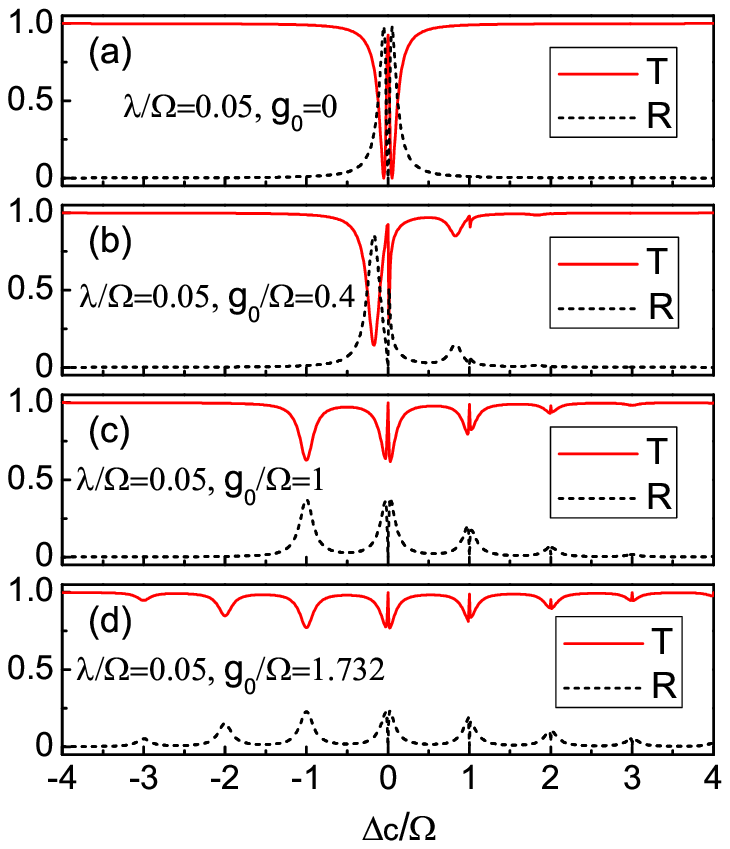}
\caption{ (Color online)
Single-photon transmission(reflection) spectra of the
atom-optomechanical system for $\lambda<\Gamma$. The parameters
are $\lambda=0.05\Omega$, $\Delta_{ac}=0$, $\gamma_{a}=0$, and
$\Gamma=0.1\Omega$.}
\label{EITlike}
\end{figure}

We first investigate the single-photon transmission and reflection
spectra of an optomechanical cavity containing no atom. Only the sideband resolved regime
$\Gamma\ll\Omega$ is considered in this
work. We plot the transmission and reflection
coefficient as the functions of the photon-cavity detuning
$\Delta_{c}$ for various values of $g_{0}$ when the mirror is
initially prepared in the ground state $\left\vert 0\right\rangle _{b}$,
as shown in Fig.~\ref{NoAtom}. When $g_{0}=0$, the coherently
interference of the leaked waves out of the cavity and the
propagating modes in the one-dimensional continuum results in
a complete suppression of the transmission for a resonantly incident
photon with $\Delta_{c}=0$. When entering the single-photon strong regime
$g_{0}>\Gamma$, the transmission dips (reflection peaks)
appear at $\Delta_{c}=-\delta+n\Omega$ ($n=0,1,2\cdots$),
exhibiting a global red shift $\delta$ and
more sidebands. This means that  an incident single-photon
with frequency $\omega_{c}-\delta+n\Omega$ can be strongly reflected by the
optomechanical system because
of the strong optomechanical coupling.

To investigate the single-photon transmission and reflection
property of the hybrid atom-optomechanical system, we first give the
eigen energies and eigen states of an isolate atom-optomechanical
system. The Hamiltonian of an isolate atom-optomechanical system can
be written as
\begin{equation}
\hat{H}_{AO}=\hat{H}_{0}+\hat{H}_{I}
\end{equation}
with
\begin{subequations}
\begin{equation}
\hat{H}_{0}=\frac{\omega_{a}}{2}\sigma_{z}+\omega_{c}c^{\dagger}c+\Omega
b^{\dagger}b-g_{0}c^{\dagger}c\left(b+b^{\dagger}\right),
\end{equation}
\begin{equation}
\hat{H}_{I}=\lambda\left(c\sigma^{+}+c^{\dagger}\sigma^{-}\right).
\end{equation}
\end{subequations}
In the single-photon subspace, the eigen states of $\hat{H}_{0}$ are
$\left\vert 0\right\rangle _{c}\left\vert e\right\rangle
_{a}\left\vert n\right\rangle _{b}$ with the eigen energy
$\epsilon_{\uparrow n}=\omega_{a}/2+n\Omega$, and $\left\vert
1\right\rangle _{c}\left\vert g\right\rangle _{a}\left\vert
\tilde{n}\right\rangle _{b}$ with the eigen energy $\epsilon_{\downarrow
n}=-\omega_{a}/2+\omega_{c}+n\Omega-\delta$. In this subspace, exact
diagonalization of the Hanmiltonian $\hat{H}_{AO}$ yields the eigen
states
\begin{subequations}
\begin{eqnarray}
\left|\psi{}_{n}^{\left(+\right)}\right\rangle=\sin\theta\left\vert 1\right\rangle _{c}\left\vert g\right\rangle _{a}\left\vert \tilde{n}\right\rangle _{b}+\cos\theta\left\vert 0\right\rangle _{c}\left\vert e\right\rangle _{a}\left\vert n\right\rangle _{b},
\\
\left|\psi{}_{n}^{\left(-\right)}\right\rangle
=-\cos\theta\left\vert 1\right\rangle _{c}\left\vert g\right\rangle
_{a}\left\vert \tilde{n}\right\rangle _{b}+\sin\theta\left\vert
0\right\rangle _{c}\left\vert e\right\rangle _{a}\left\vert
n\right\rangle _{b}
\end{eqnarray}
\end{subequations}
with the corresponding eigen energies
\begin{subequations}
\begin{eqnarray}
E_{n}^{\left(+\right)}&=&\frac{\omega_{c}}{2}+n\Omega-\frac{\delta}{2}+\frac{1}{2}\sqrt{\left(\Delta_{ac}+\delta\right)^{2}+4\lambda^{2}},
\\
E_{n}^{\left(-\right)}&=&\frac{\omega_{c}}{2}+n\Omega-\frac{\delta}{2}-\frac{1}{2}\sqrt{\left(\Delta_{ac}+\delta\right)^{2}+4\lambda^{2}},
\end{eqnarray}
\end{subequations}
where
\[
\theta=\frac{1}{2}\tan^{-1}\frac{2\lambda}{\Delta_{ac}+\delta}.
\]

We now consider the case that the two-level atom is in resonance
with the cavity, i.e. $\Delta_{ac}=0$. When the atom-cavity coupling
strength $\lambda\gg\Omega\gg\Gamma$ and the optmechanical coupling strength
$g_{0}=0$, the transmission(reflection) spectrum shows vacuum Rabi
splitting~\cite{cQED1,Shen2009} with the splitting width $2\lambda$, as
shown in Fig.~\ref{Rabilike}(a). Figs.~\ref{Rabilike}(b)-(d) show
that  how the moving
mirror modify vacuum Rabi spectrum in the single-photon strong coupling regime. When the coupling strength $g_{0}$ increases to
enter into the single-photon strong coupling regime, the spectra will
undergo a red shift $\delta/2$. Additionally, on the right side of
each main peak, more sidebands will appear with interval $\Omega$,
corresponding to the energy levels (under the condition
$\lambda\gg\delta$)
\begin{subequations}
\begin{eqnarray}
E_{n}^{\left(+\right)}\thickapprox\frac{\omega_{c}}{2}+n\Omega-\frac{\delta}{2}+\lambda,
\\
E_{n}^{\left(-\right)}\thickapprox\frac{\omega_{c}}{2}+n\Omega-\frac{\delta}{2}-\lambda.
\end{eqnarray}
\end{subequations}
The corresponding eigen states take the form
\begin{subequations}
\begin{eqnarray}
\left|\psi{}_{n}^{\left(+\right)}\right\rangle \thicksim\frac{1}{\sqrt{2}}\left(\left\vert 1\right\rangle _{c}\left\vert g\right\rangle _{a}\left\vert \tilde{n}\right\rangle _{b}+\left\vert 0\right\rangle _{c}\left\vert e\right\rangle _{a}\left\vert n\right\rangle _{b}\right),
\\
\left|\psi{}_{n}^{\left(-\right)}\right\rangle
\thicksim\frac{1}{\sqrt{2}}\left(\left\vert 1\right\rangle
_{c}\left\vert g\right\rangle _{a}\left\vert \tilde{n}\right\rangle
_{b}-\left\vert 0\right\rangle _{c}\left\vert e\right\rangle
_{a}\left\vert n\right\rangle _{b}\right).
\end{eqnarray}
\end{subequations}
The energy-level structure in this case is potted in Fig.~\ref{EL}(a).

If $\lambda<\Gamma$, we can get a spectrum that is analogous to that for
electromagnetically induced transparency (EIT)
phenomena~\cite{Shen2009,EIT}. Typically, when $g_{0}=0$, the
spectrum exhibits a standard EIT one with a very narrow transmission window,
as shown in Fig.~\ref{EITlike} (a). When entering the single-photon
strong coupling regime $g_{0}>\Gamma$, more EIT structures appear in
the sideband regime (Figs. ~\ref{EITlike}(b)-(d)). The transmission
maxima are located at $\Delta_{c}=n\Omega$ ($n=0,1,2\cdots$).
Typically, when $g_{0}=\sqrt{m}\Omega$ ($m=1,2,\cdots$), i.e.,
$\delta=m\Omega$, we have $\epsilon_{\uparrow
n}=\epsilon_{\downarrow n+m}=\omega_{c}/2+n\Omega$. Namely,
the eigen states of $H_{0},$ $\left\vert 0\right\rangle
_{c}\left\vert e\right\rangle _{a}\left\vert n\right\rangle _{b}$
and $\left\vert 1\right\rangle _{c}\left\vert g\right\rangle
_{a}\left\vert \tilde{n+m}\right\rangle _{b}$ are degenerate. This
degeneracy is perturbed by the relatively weak atom-cavity
interaction $H_{I}$, resulting in a pair of near degenerate states
\begin{subequations}
\begin{eqnarray}
\left|\psi{}_{n}^{\left(+\right)}\right\rangle \thicksim\left\vert 0\right\rangle _{c}\left\vert e\right\rangle _{a}\left\vert n\right\rangle _{b}+\frac{\lambda}{\delta}\left\vert 1\right\rangle _{c}\left\vert g\right\rangle _{a}\left\vert \tilde{n}\right\rangle _{b},
\\
\left|\psi{}_{n+m}^{\left(-\right)}\right\rangle \thicksim\left\vert
1\right\rangle _{c}\left\vert g\right\rangle _{a}\left\vert
\tilde{n+m}\right\rangle _{b}-\frac{\lambda}{\delta}\left\vert
0\right\rangle _{c}\left\vert e\right\rangle _{a}\left\vert
n+m\right\rangle _{b}
\end{eqnarray}
\end{subequations}
with the eigen energies
\begin{subequations}
\begin{eqnarray}
E_{n}^{\left(+\right)}\approx\frac{\omega_{c}}{2}+n\Omega+\frac{\lambda^{2}}{\delta},
\\
E_{n+m}^{\left(-\right)}\approx\frac{\omega_{c}}{2}+n\Omega-\frac{\lambda^{2}}{\delta}.
\end{eqnarray}
\end{subequations}
The energy-level structure of this case is depicted in Fig.~\ref{EL}{(b) by choosing $m=1$ as an example. 
Thus, when a single-photon with detuning $\Delta_{c}=n\Omega$ ($n=0,1,2\cdots$)
injected, destructive quantum interference occurs between the two
possible transition channel $\left\vert 0\right\rangle
_{c}\left\vert g\right\rangle _{a}\left\vert 0\right\rangle
_{b}\rightarrow\left|\psi{}_{n}^{\left(+\right)}\right\rangle $ and
$\left\vert 0\right\rangle _{c}\left\vert g\right\rangle
_{a}\left\vert 0\right\rangle
_{b}\rightarrow\left|\psi{}_{n+m}^{\left(-\right)}\right\rangle $,
resulting in a complete transmission of the single-photon. This generate a
EIT-like structure at
$\Delta_{c}=n\Omega$ in the transmission(reflection) spectrum, as 
seen in Figs.~\ref{EITlike}(c) and (d). In
addition, there are single transmission dips (reflection peaks) located at
$\Delta_{c}=-l\Omega$ ($l=1,\cdots,m$), corresponding to the
$\left\vert 0\right\rangle _{c}\left\vert g\right\rangle
_{a}\left\vert 0\right\rangle
_{b}\rightarrow\left|\psi{}_{m-l}^{\left(-\right)}\right\rangle$
($l=1,\cdots,m$) transition. 

\subsection{\label{B}Single-photon transmission(reflection) coefficient:
effects of atom-cavity detunings and dissipations}

\begin{figure}[t]
\includegraphics[width=0.5\textwidth]{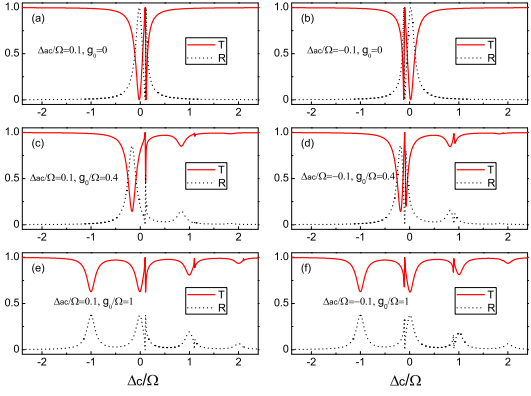}
\caption{ (Color online)
Single-photon transmission(reflection) spectra for detuned atom-cavity
 cases. In (a), (c), and (e), the atom-cavity detuning is
$\Delta_{ac}=-0.1\Omega$; in (b), (d), and (f),
$\Delta_{ac}=0.1\Omega$. The other parameters are
$\lambda=0.05\Omega$, $\gamma_{a}=0$, and $\Gamma=0.1\Omega$.}
\label{detuning}
\end{figure}

\begin{figure}[b]
\includegraphics[width=0.5\textwidth]{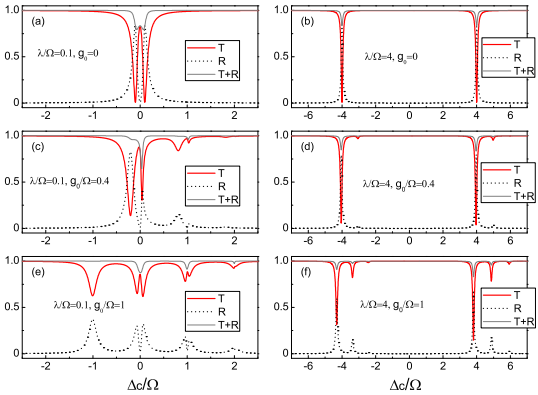}
\caption{(Color online) Single-photon transmission (reflection)
spectra for dissipative atom cases ($\gamma_{a}=0,01\Omega$). In
(a), (c), and (e), the atom-cavity coupling strength is
$\lambda=0.1\Omega$; in (b), (d), and (f), $\lambda=4\Omega$. The
other parameters are $\Delta_{ac}=0$ and $\Gamma=0.1\Omega$.}
\label{dissipation}
\end{figure}

We next consider the case of atom cavity to be detuned. When the
optomechanical coupling strength $g_{0}=0$, for a photon on resonance
with the atom $\Delta_{c}=\Delta_{ac}$, the transmission amplitude
is always 1, as seen from Figs.~\ref{detuning} (a) and (b), 
which was indicated in Ref.~\cite{Shen2009}. When entering the
single-photon strong coupling regime $g>\Gamma$, we can see that
these maxima will appear at $\Delta_{c}=\Delta_{ac}+n\Omega$,
corresponding to the $\left\vert 0\right\rangle _{c}\left\vert
g\right\rangle _{a}\left\vert 0\right\rangle
_{b}\rightarrow\left\vert 0\right\rangle _{c}\left\vert
e\right\rangle _{a}\left\vert n\right\rangle _{b}$ transitions, as
shown in Figs.~\ref{detuning} (c)-(f).

There exist unavoidable intrinsic dissipative processes in the
system, resulting in the leakage of photons into non-waveguided
degrees of freedom. Here we assume that the cavity strongly coupled
to the waveguide, namely, the majority of the decayed light from the
cavity is guided into waveguide modes~\cite{HXZheng2011}. Thus the
decay rate $\kappa$ of the cavity into channels other than the 1D
continuum is negligible. The main dissipative processes are originated from
the decay of atom. Experimentally, in both typical cavity QED 
and solid-state circuit QED systems, the ratio between the atom
(artificial atom) decay rate and the cavity decay rate is about
$\gamma_{a}/\Gamma\sim0.1$~\cite{Blais2004}. Figs.~\ref{dissipation}
give the transmission(reflection) spectrum of the dissipative atom
case. The leakage of photons into non-waveguided degrees of freedom
can be measured in terms of $T+R$ (the grey thin lines). In the Rabi-splitting like cases with
strong atom-cavity coupling ($\lambda\gg\Omega\gg\Gamma$), the atom
dissipation has a stronger effect on the transmission of an photon
at the frequencies of resonant absorption. On the contrary, in the EIT-like cases with
relatively small atom-cavity coupling strength
($\lambda\sim\Gamma$), the atom dissipation has a stronger effect on
the transmission of a photon with detuning around
$\Delta_{c}=n\Omega$ ($n=0,1,2,\cdots$).

\subsection{\label{C}The final reservoir occupation spectrum}

\begin{figure}[t]
\includegraphics[width=0.5\textwidth]{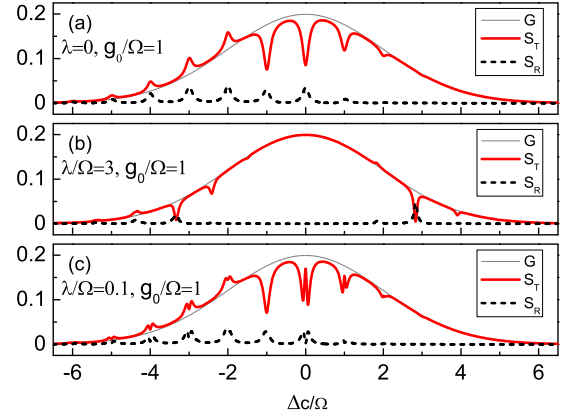}
\caption{(Color online) Transmitted (reflected)
photon spectra $S_{T}\left(\Delta_{c}\right)$
($S_{R}\left(\Delta_{c}\right)$ ). The parameters
are $\Delta_{0}=0$, $d=4\Omega$, $\Delta_{ac}=0$, $\gamma_{a}=0,01\Omega$, and
$\Gamma=0.1\Omega$. The grey thin curve is the spectral density $G$ of incident photons.}
\label{occupation spectrum}
\end{figure}

 We have above discussed the transmission (reflection) coefficients of
a monochromatic incident photon. Note that for an optomechanical
system in single-photon strong coupling regime, the
inelastic scattering should have an influential effect, resulting in a re-emitted
photon with red/blue sideband frequency. This is different from the
case of photon transmitting (reflecting) from a usual cavity, where
the frequency of the photon remains unchanged after scattering. To
see this point more clearly, we calculate the final reservoir occupation
spectrum~\cite{Liao2012,Liao2013,Xiao2013}, which describes
probability density for finding the single photon with a specific
frequency of the transmission (reflection) fields. Let us consider an
incident photon with a Gaussian-type spectral amplitude
$\alpha(\Delta_{c})=\left(2/\pi
d^{2}\right)^{1/4}\exp[-\left(\Delta_{c}-\Delta_{0}\right)^{2}/d^{2}]$
, where $\Delta_{0}$ and $d$ is the detuning and spectrum width of
the photon, respectively. The according spectral density can be
represented as $G={\left|\alpha\right|}^{2}$. We plot in
Fig.~\ref{occupation spectrum} the spectra
$S_{T}\left(\Delta_{c}\right)$ and $S_{R}\left(\Delta_{c}\right)$ of
resonantly incident single-photon (i.e., $\Delta_{0}=0$) scattering
when the mirror is initially prepared in the ground state $\left\vert
0\right\rangle _{b}$, with strong optomechanical coupling strength
$g_{0}=\Omega$ and different atom-cavity coupling strength
$\lambda$. It can be seen from Fig.~\ref{occupation spectrum} that
phonon sidebands appear in the spectrum in the single-photon strong
coupling regime. The dips in the spectrum
$S_{T}\left(\Delta_{c}\right)$ appear at the same position  as in
the spectrum $T$, correspond to the resonant transition from
$\left\vert 0\right\rangle _{c}\left\vert 0\right\rangle
_{a}\left\vert 0\right\rangle _{b}$ to the excited states. Thus after
scattering, the probability density for finding the single photon at
these frequencies decreases. And the peaks in the red sideband indicate
that the re-emitted photon can lose its energy by $n\Omega$,
leading to the final state $\left\vert 0\right\rangle _{c}\left\vert
0\right\rangle _{a}\left\vert n\right\rangle _{b}$ of system and
increasing the value of $S_{T}\left(\Delta_{c}\right)$ at these
transition frequencies.

\section{\label{4}Conclusions}

In summary, we have for the first time explored the single-photon transport in a
waveguide coupled to a hybrid atom-optomechanical system in the
single-photon strong-coupling regime. These spectra can characterize
the mirror-cavity and atom-cavity couplings. On one hand, an
optomechanical coupling dependent frequency shift and more sidebands
appear in the transmission (reflection) spectra when the optical
coupling strength increases. For the existence of atomic degrees of
freedom, we can get a Rabi-splitting like or an EIT-like spectrum,
depending on the atom-cavity coupling strength. Here we wish to make some
further remarks on the possible experimental realizations of hybrid
atom-optomechanical systems. Such a hybrid system can be possibly
realized by directly combining the well developed technology of
optomechanical cavities with moving mirror\cite{Kimble2005} and
trapping a single atom in cavity
QED~\cite{Kippenberg2008,Marquardt2009,Aspelmeyer2013}. Also, this
set up may be more easily achieved using an on-chip circuit cavity
electromechanics with a spiral inductor shunted by a parallel-plate
capacitor, an analog of optomechanical cavity in microwave
domain~\cite{ExpStrongCoup4}, which can be easily coupled to a
superconducting artificial atom using currently availabe circuit QED
technology~\cite{cQED1}. These systems may provide quantum interface
allowing the coherent transfer of quantum states between the
mechanical oscillator and atoms, opening a door for coherent
preparation and manipulation of micromechanical
resonators~\cite{OMwithAtom1}.

\begin{acknowledgments}
We thank helpful discussions with Y. Li. This work was supported by the GRF (HKU7058/11P) and the CRF (HKU8/11G) of Hong Kong,
and the URC fund of HKU.
\end{acknowledgments}






\begin{thebibliography}{99}
\bibitem{Shen2005} J. T. Shen and S. Fan, Phys. Rev. Lett., \textbf{95},
213001 (2005).

\bibitem{Shen2009} J. T. Shen and S. Fan, Phys. Rev. A \textbf{79},
023837 (2009).

\bibitem{Zhou2008} L. Zhou, Z. R. Gong, Y. X. Liu, C. P. Sun, and
F. Nori, Phys. Rev. Lett. \textbf{101}, 100501 (2008).

\bibitem{photonic} J. Claudon, J. Bleuse, N. S. Malik, M. Bazin,
P. Jaffrennou, N. Gregersen, C. Sauvan, P. Lalanne and Jean-Michel
G�rard, Nat. Photon. \textbf{4}, 174 (2010).

\bibitem{plasmon} D. E. Chang, A. S. S�rensen, E. A. Demler, and
M. D. Lukin, Nature Phys. \textbf{3}, 807 (2007).

\bibitem{NECscience2010} O. Astafiev, A. M. Zagoskin, A. A. Abdumalikov
Jr., Y. A. Pashkin, T. Yamamoto, K. Inomata, Y. Nakamura and J. S.
Tsai, Science \textbf{327}, 840 (2010).

\bibitem{cQED1} A. Wallraff, D. I. Schuster, A. Blais, L. Frunzio,
R. S. Huang, J. Majer, S. Kumar, S. M. Girvin, and R. J. Schoelkopf,
Nature (London) \textbf{431}, 162 (2004);

\bibitem{cQED2} T. Aoki, B. Dayan, E. Wilcut, W. P. Bowen, A. S.
Parkins, T. J. Kippenberg, K. J. Vahala, and H. J. Kimble, Nature
(London) \textbf{443}, 671 (2006).

\bibitem{cQED3} K. Srinivasan and O. Painter, Nature (London) \textbf{450},
862 (2007).

\bibitem{cQED4} B. Dayan, A. S. Parkins, T. Aoki, E. P. Ostby, K.
J. Vahala, and H. J. Kimble, Science \textbf{319}, 1062 (2008).

\bibitem{Kippenberg2008} T. J. Kippenberg and K. J. Vahala, Science
\textbf{321}, 1172 (2008).

\bibitem{Marquardt2009} F. Marquardt and S. M. Girvin, Physics \textbf{2},
40 (2009).

\bibitem{Aspelmeyer2013} M. Aspelmeyer, T. J. Kippenberg, and F.
Marquardt, arXiv:1303.0733.

\bibitem{ExpStrongCoup1} K. W. Murch, K. L. Moore, S. Gupta, and
D. M. Stamper-Kurn, Nat Phys \textbf{4}, 561 (2008).

\bibitem{ExpStrongCoup2} F. Brennecke, S. Ritter, T. Donner, and
T. Esslinger, Science \textbf{322}, 235 (2008).

\bibitem{ExpStrongCoup3} D. W. C. Brooks, T. Botter, S. Schreppler,
T. P. Purdy, N. Brahms, and D. M. Stamper-Kurn, Nature \textbf{488},
476 (2012).

\bibitem{ExpStrongCoup4} J. D. Teufel, T. Donner, D. Li, J. W. Harlow,
M. S. Allman, K. Cicak, A. J. Sirois, J. D. Whittaker, K. W.
Lehnert, and R. W. Simmonds, Nature \textbf{475}, 359 (2011).

\bibitem{ExpStrongCoup5} J. Chan, T. P. M. Alegre, A. H. Safavi-Naeini,
J. T. Hill, A. Krause, S. Groblacher, M. Aspelmeyer, and O. Painter,
Nature \textbf{478}, 89 (2011).

\bibitem{ExpStrongCoup6} E. Verhagen, S. Deleglise, S. Weis, A. Schliesser,
and T. J. Kippenberg, Nature \textbf{482}, 63 (2012).

\bibitem{ExpStrongCoup7} J. Chan, A. H. Safavi-Naeini, J. T. Hill,
S. Meenehan, and O. Painter, Applied Physics Letters \textbf{101},
081115 (2012).

\bibitem{Rabl2011 }P. Rabl, Phys. Rev. Lett. \textbf{10}7, 063601
(2011).

\bibitem{Nunnenkamp2011}A. Nunnenkamp, K. B�rkje, and S. M. Girvin,
Phys. Rev. Lett. \textbf{10}7, 063602 (2011).

\bibitem{Liao2013}Jie-Qiao Liao and C. K. Law, Phys. Rev. A \textbf{8}7,
043809 (2013).

\bibitem{Liu2013}Xun-Wei Xu, Yuan-Jie Li, and Yu-xi Liu, Phys. Rev.
A \textbf{87}, 025803 (2013).

\bibitem{SPcooling}A. Nunnenkamp, K. B�rkje, and S. M. Girvin, Phys.
Rev. A \textbf{85}, 051803 (2012).

\bibitem{OMIT}A. Kronwald and F. Marquardt, arXiv:1304.5230.

\bibitem{OM instability}J. Qian, A. A. Clerk, K. Hammerer, and F.
Marquardt, Phys. Rev. Lett. \textbf{109}, 253601 (2012).

\bibitem{Liao2012}J. Q. Liao, H. K. Cheung, and C. K. Law, Phys.
Rev. A \textbf{85}, 025803 (2012).

\bibitem{Xiao2013}Xue-Xin Ren, Hao-Kun Li, Meng-Yuan Yan, Yong-Chun
Liu, Yun-Feng Xiao, and Qihuang Gong, Phys. Rev. A \textbf{87},
033807 (2013).

\bibitem{OMwithAtom1} K. Hammerer, M. Wallquist, C. Genes, M. Ludwig,
F. Marquardt, P. Treutlein, P. Zoller, J. Ye, and H. J. Kimble,
Phys. Rev. Lett. \textbf{103}, 063005 (2009); M. Wallquist, K.
Hammerer, P. Zoller, C. Genes, M. Ludwig, F. Marquardt, P.
Treutlein, J. Ye, and H. J. Kimble, Phys. Rev. A \textbf{81}, 023816
(2010).

\bibitem{OMwithAtom2} Daniel Breyer and Marc Bienert, Phys. Rev.
A \textbf{86}, 053819 (2012).

\bibitem{Kimble2005} R. Miller, T. E. Northup, K. M. Birnbaum, A.
Boca, A. D. Boozer, and H. J. Kimble, J. Phys. B \textbf{38}, S551
(2005).

\bibitem{HXZheng2011} Huaixiu Zheng, D. J. Gauthier, and H. U. Baranger,
Phys. Rev. lett. \textbf{107}, 223601 (2011).

\bibitem{Blais2004} A. Blais, R. S. Huang, A. Wallraff, S. M. Girvin,
and R. J. Schoelkopf Phys. Rev. A \textbf{69}, 062320 (2004).

\bibitem{EIT}S. E. Harris, Phys. Today \textbf{50}(7), 36 (1997).
\end{thebibliography}
\end{document}